\documentclass[letterpaper,aps,prl,superscriptaddress,showpacs,floatfix,twocolumn]{revtex4}

\usepackage{graphicx}
\usepackage{hyperref}
\usepackage{amsmath}
\usepackage{url}

\usepackage{times}

\begin{document}

\title{ Flavor Asymmetry of the Nucleon Sea and the Five-Quark
Components of the Nucleons}

\author{Wen-Chen Chang}
\affiliation{Institute of Physics, Academia Sinica, Taipei 11529, Taiwan}

\author{Jen-Chieh Peng}
\affiliation{Institute of Physics, Academia Sinica, Taipei 11529, Taiwan}
\affiliation{Department of Physics, University of Illinois at 
Urbana-Champaign, Urbana, Illinois 61801, USA}

\date{\today}

\begin{abstract}
The existence of the five-quark Fock states for the intrinsic charm
quark in the nucleons was suggested some time ago, but conclusive
evidence is still lacking. We generalize the previous theoretical
approach to the light-quark sector and study possible experimental
signatures for such five-quark states. In particular, we compare the
$\bar d - \bar u$ and $\bar u + \bar d - s -\bar s$ data with the
calculations based on the five-quark Fock states. The qualitative
agreement between the data and the calculations is interpreted as
evidence for the existence of the intrinsic light-quark sea in the
nucleons. The probabilities for the $|uudu\bar{u}\rangle$ and
$|uudd\bar{d}\rangle$ Fock states are also extracted.
\end{abstract}

\pacs{12.38.Lg,14.20.Dh,14.65.Bt,14.65.Dw,13.60.Hb}

\maketitle


The possible existence of a significant $u u d c \bar c$ five-quark
Fock component in the proton was proposed some time ago by Brodsky,
Hoyer, Peterson, and Sakai (BHPS)~\cite{brodsky80} to explain the
unexpectedly large production rates of charmed hadrons at large
forward $x_F$ region. In the light-cone Fock space framework, the
probability distribution of the momentum fraction (Bjorken-$x$) for
this nonperturbative ``intrinsic" charm (IC) component was
obtained~\cite{brodsky80}. The intrinsic charm originating from the
five-quark Fock state is to be distinguished from the ``extrinsic"
charm produced in the splitting of gluons into $c \bar c$ pairs, which
is well described by QCD. The extrinsic charm has a ``sea-like"
characteristics with large magnitude only at the small $x$ region. In
contrast, the intrinsic charm is ``valence-like" with a distribution
peaking at larger $x$. The presence of the intrinsic charm component
can lead to a sizable charm production at the forward rapidity ($x_F$)
region.

The $x$ distribution of the intrinsic charm in the BHPS model was
derived with some simplifying assumptions. Recently,
Pumplin~\cite{pumplin06} showed that a variety of light-cone models in
which these assumptions are removed would still predict the $x$
distributions of the intrinsic charm similar to that of the BHPS
model. The CTEQ collaboration~\cite{pumplin06} has also examined all
relevant hard-scattering data sensitive to the presence of the IC
component, and concluded that the existing data are consistent with a
wide range of the IC magnitude, from null to 2-3 times larger than the
estimate by the BHPS model. This result shows that the experimental
data are not yet sufficiently accurate to determine the magnitude or
the $x$ distribution of the IC.

In an attempt to further study the role of five-quark Fock states for
intrinsic quark distributions in the nucleons, we have extended the
BHPS model to the light quark sector and compared the predictions with
the experimental data. The BHPS model predicts the probability for the
$u u d Q \bar Q$ five-quark Fock state to be approximately
proportional to $1/m_Q^2$, where $m_Q$ is the mass of the quark
$Q$~\cite{brodsky80}. Therefore, the light five-quark states $u u d u
\bar u$ and $u u d d \bar d$ are expected to have significantly larger
probabilities than the $u u d c \bar c$ state. This suggests that the
light quark sector could potentially provide more clear evidence for
the roles of the five-quark Fock states, allowing the specific
predictions of the BHPS model, such as the shape of the quark $x$
distributions originating from the five-quark configuration, to be
tested.

To compare the experimental data with the prediction based on the
intrinsic five-quark Fock state, it is essential to separate the
contributions of the intrinsic quark and the extrinsic
one. Fortunately, there exist some experimental observables which are
free from the contributions of the extrinsic quarks. As discussed
later, the $\bar d - \bar u$ and the $\bar u + \bar d - s - \bar s$
are examples of quantities independent of the contributions from
extrinsic quarks. The $x$ distribution of $\bar d - \bar u$ has been
measured in a Drell-Yan experiment~\cite{e866}. A recent measurement
of $s + \bar s$ in a semi-inclusive deep-inelastic scattering (DIS)
experiment~\cite{hermes} also allowed the determination of the $x$
distribution of $\bar u + \bar d - s - \bar s$. In this paper, we
compare these data with the calculations based on the intrinsic
five-quark Fock states. The qualitative agreement between the data and
the calculations provides evidence for the existence of the intrinsic
light-quark sea in the nucleons.


For a $|u u d Q \bar Q\rangle$ proton Fock state, the probability for
quark $i$ to carry a momentum fraction $x_i$ is given in the BHPS
model~\cite{brodsky80} as
\begin{equation}
P(x_1, ...,x_5)=N_5\delta(1-\sum_{i=1}^5x_i)[m_p^2-\sum_{i=1}^5\frac{m_i^2}{x_i}]^{-2},
\label{eq:prob5q_a}
\end{equation}
\noindent where the delta function ensures momentum
conservation. $N_5$ is the normalization factor for five-quark Fock
state, and $m_i$ is the mass of quark $i$. In the limit of $m_{4,5} >>
m_p, m_{1,2,3}$, where $m_p$ is the proton mass, Eq.~\ref{eq:prob5q_a}
becomes
\begin{equation}
P(x_1, ...,x_5)=\tilde{N}_5\frac{x_4^2x_5^2}{(x_4+x_5)^2} \delta(1-\sum_{i=1}^5 x_i),
\label{eq:prob5q_b}
\end{equation}
\noindent where $\tilde{N}_5 = N_5/m_{4,5}^4$. Eq.~\ref{eq:prob5q_b}
can be readily integrated over $x_1$, $x_2$, $x_3$ and $x_4$, and the
heavy-quark $x$ distribution~\cite{brodsky80,pumplin06} is:
\begin{eqnarray}
P(x_5)=\frac{1}{2} \tilde{N}_5 x_5^2[\frac{1}{3} (1-x_5)
(1+10x_5+x_5^2) \nonumber \\
-2x_5(1+x_5)\ln (1/x_5)].
\label{eq:prob5q_d}
\end{eqnarray}
\noindent One can integrate Eq.~\ref{eq:prob5q_d} over $x_5$ and
obtain the result ${\cal P}^{c \bar c}_5 = \tilde{N}_5/3600$, where
${\cal P}^{c \bar c}_5$ is the probability for the $|u u d c \bar
c\rangle$ five-quark Fock state. An estimate of the magnitude of
${\cal P}^{c \bar c}_5$ was given by Brodsky et al.~\cite{brodsky80}
as $\approx 0.01$, based on diffractive production of
$\Lambda_c$. This value is consistent with a bag-model
estimate~\cite{donoghue77}.

\begin{figure}[t]
\includegraphics[width=0.5\textwidth]{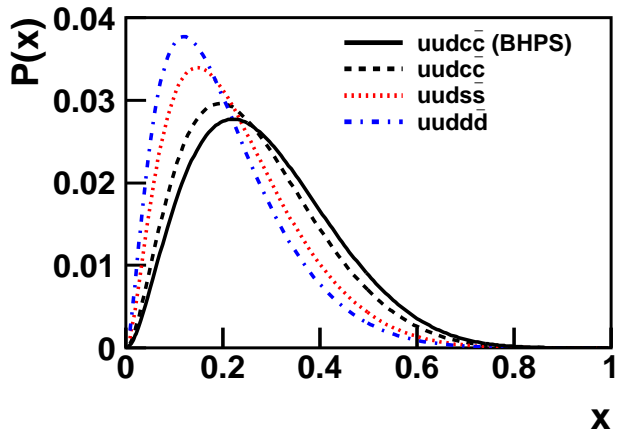}
\caption{The $x$ distributions of the intrinsic $\bar Q$ in the $u u d
Q \bar Q$ configuration of the proton from the BHPS
model~\cite{brodsky80}. The solid curve is plotted using the
expression in Eq.~\ref{eq:prob5q_d} for $\bar c$. The other three
curves, corresponding to $\bar c$, $\bar s$, and $\bar d$ in the
five-quark configurations, are obtained by solving
Eq.~\ref{eq:prob5q_a} numerically. The same probability ${\cal P}^{Q
\bar Q}_5$ (${\cal P}^{Q \bar Q}_5= 0.01$) is used for the three
different five-quark states.}
\label{fig_5q_c_s_d}
\end{figure}

The solid curve in Fig.~\ref{fig_5q_c_s_d} shows the $x$ distribution
for the charm quark ($P(x_5)$) using Eq.~\ref{eq:prob5q_d}, assuming
${\cal P}^{c \bar c}_5 = 0.01$. Since this analytical expression was
obtained for the limiting case of infinite charm-quark mass, it is of
interest to compare this result with calculations without such an
assumption. To this end, we have developed the algorithm to calculate
the quark distributions using Eq.~\ref{eq:prob5q_a} with Monte-Carlo
techniques. The five-quark configuration of $\{x_1,...,x_5\}$
satisfying the constraint of Eq.~\ref{eq:prob5q_a} is randomly
sampled. The probability distribution $P(x_i)$ can be obtained
numerically with an accumulation of sufficient statistics. We first
verified that the Monte-Carlo calculations in the limit of very heavy
charm quarks reproduce the analytical result for $P(x_5)$ in
Eq.~\ref{eq:prob5q_d}. We then calculated $P(x_5)$ using $m_u = m_d =
0.3$ GeV/$c^2$, $m_c = 1.5$ GeV/$c^2$, and $m_p = 0.938$ GeV/$c^2$,
and the result is shown as the dashed curve in
Fig.~\ref{fig_5q_c_s_d}. The similarity between the solid and dashed
curves shows that the assumption adopted for deriving
Eq.~\ref{eq:prob5q_d} is adequate. It is important to note that the
Monte-Carlo technique allows us to calculate the quark $x$
distributions for other five-quark configurations when $Q$ is the
lighter $u$, $d$, or $s$ quark, for which one could no longer assume a
large mass.

As mentioned above, the insufficient accuracy of existing data as well
as the inherently small probability for intrinsic charm due to the
large charm-quark mass make it difficult to confirm the existence of
the intrinsic charm component in the proton. On the other hand the
five-quark states involving only lighter quarks, such as $|u u d u
\bar u\rangle$, $|u u d d \bar d\rangle$, and $|u u d s \bar
s\rangle$, might be more easily observed experimentally. We have
calculated the $x$ distributions of the $\bar s$ and $\bar d$ quarks
in the BHPS model for the $|u u d s \bar s\rangle$ and $|u u d d \bar
d\rangle$ configurations, respectively, using
Eq.~\ref{eq:prob5q_a}. The mass of the strange quark is chosen as 0.5
GeV/c$^2$. In Fig.~\ref{fig_5q_c_s_d}, we show the $x$ distributions
of $\bar s$ and $\bar d$, together with that of $\bar c$. In order to
focus on the different shapes of the $x$ distributions, the same value
of ${\cal P}^{Q \bar Q}_5$ is assumed for these different five-quark
states. Figure~\ref{fig_5q_c_s_d} shows that the $x$ distributions of
the intrinsic $\bar Q$ shift progressively to lower $x$ region as the
mass of the quark $Q$ decreases. The $x$ distributions of $\bar Q$
originating from the gluon splitting into quark-antiquark pair ($g \to
Q \bar Q$) QCD processes are localized at the low-$x$
region. Figure~\ref{fig_5q_c_s_d} illustrates an important advantage
for identifying the IC component, namely, the intrinsic charm
component is better separated from the extrinsic charm component as a
result of their different $x$ distributions. Nevertheless, the
probability for intrinsic lighter quarks are expected to be
significantly larger than for the heavier charm quark. The challenge
is to identify proper experimental observables which allow a clear
separation of the intrinsic light quark component from the extrinsic
QCD component. As we discuss next, the quantities $\bar d(x) - \bar
u(x)$ and $\bar u(x) + \bar d(x) - s(x) - \bar s(x)$ are suitable for
studying the intrinsic light-quark components of the proton.


\begin{figure}[b]
\includegraphics[width=0.5\textwidth]{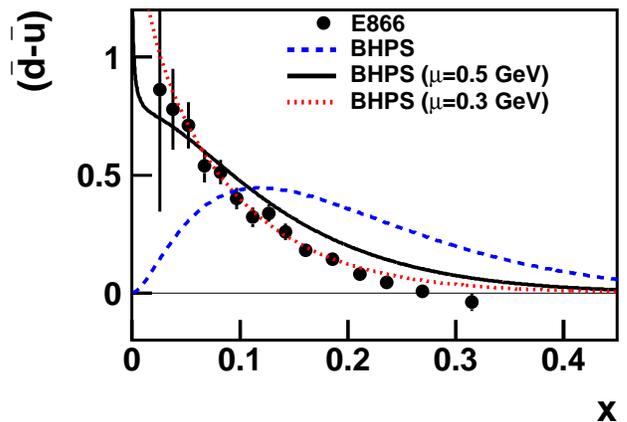}
\caption{Comparison of the $\bar d(x) - \bar u(x)$ data with the
calculations based on the BHPS model. The dashed curve corresponds to
the calculation using Eq.~\ref{eq:prob5q_a} and
Eq.~\ref{eq:intdbarubar2}, and the solid and dotted curves are
obtained by evolving the BHPS result to $Q^2 = 54.0$ GeV$^2$ using
$\mu = 0.5$ GeV and $\mu = 0.3$ GeV, respectively.}
\label{fig_dbar-ubar}
\end{figure}

The first evidence for an asymmetric $\bar u$ and $\bar d$
distribution came from the observation~\cite{amaudruz91} that the
Gottfried Sum Rule~\cite{gottfried67} was violated. The striking
difference between the $\bar d$ and $\bar u$ distributions was clearly
observed subsequently in the proton-induced Drell-Yan~\cite{na51,e866}
and semi-inclusive DIS experiments~\cite{hermes_sidis}. This large
flavor asymmetry was in qualitative agreement with the meson cloud
model which incorporates chiral symmetry~\cite{MCM1}. Reviews on this
subject can be found in Refs.~\cite{tony,kumano,garvey}.

The $\bar d(x) - \bar u(x)$ data from the Fermilab E866 Drell-Yan
experiment at the $Q^2$ scale of 54 GeV$^2$~\cite{e866} is shown in
Fig.~\ref{fig_dbar-ubar}. The $\bar d(x) - \bar u(x)$ distribution is
of particular interest for testing the intrinsic light-quark contents
in the proton, since the perturbative $g \to Q \bar Q$ processes are
expected to generate $u \bar u$ and $d \bar d$ pairs with equal
probabilities and thus have no contribution to this quantity. In the
BHPS model, the $\bar u$ and $\bar d$ are predicted to have the same
$x$ dependence if $m_u = m_d$. It is important to note that the
probabilities of the $|u u d d \bar d\rangle$ and $|u u d u \bar
u\rangle$ configurations, ${\cal P}^{u \bar u}_5$ and ${\cal P}^{d
\bar d}_5$, are not known from the BHPS model, and remain to be
determined from the experiments. Non-perturbative effects such as
Pauli-blocking~\cite{feynman} could lead to different probabilities
for the $|u u d d \bar d\rangle$ and $|u u d u \bar u\rangle$
configurations. Nevertheless the shape of the $\bar d(x) - \bar u(x)$
distribution shall be identical to those of $\bar d(x)$ and $\bar
u(x)$ in the BHPS model. Moreover, the normalization of $\bar d(x) -
\bar u(x)$ is already known from the Fermilab E866 Drell-Yan
experiment as
\begin{equation}
\int^{1}_{0} (\bar d(x) - \bar u(x)) dx = 0.118 \pm 0.012 .
\label{eq:intdbarubar1}
\end{equation}
\noindent This allows us to compare the $\bar d(x) - \bar u(x)$ data
with the calculations from the BHPS model, since the above integral is
simply equal to ${\cal P}^{d \bar d}_5 - {\cal P}^{u \bar u}_5$, i.e.
\begin{equation}
\int^{1}_{0} (\bar d(x) - \bar u(x)) dx = {\cal P}^{d \bar d}_5 -
{\cal P}^{u \bar u}_5 = 0.118 \pm 0.012 .
\label{eq:intdbarubar2}
\end{equation}

Figure~\ref{fig_dbar-ubar} shows the calculation of the $\bar d(x) -
\bar u(x)$ distribution (dashed curve) from the BHPS model, together
with the data. The $x$-dependence of the $\bar d(x) - \bar u(x)$ data
is not in good agreement with the calculation. It is important to note
that the $\bar d(x) - \bar u(x)$ data in Fig.~\ref{fig_dbar-ubar} were
obtained at a rather large $Q^2$ of 54 GeV$^2$~\cite{e866}. In
contrast, the relevant scale, $\mu^2$, for the five-quark Fock states
is expected to be much lower, around the confinement scale. This
suggests that the apparent discrepancy between the data and the BHPS
model calculation in Fig.~\ref{fig_dbar-ubar} could be partially due
to the scale dependence of $\bar d(x) - \bar u(x)$. We adopt the value
of $\mu = 0.5$ GeV, which was chosen by Gl\"{u}ck, Reya, and
Vogt~\cite{grv} in their attempt to generate gluon and quark
distributions in the so-called ``dynamical approach" starting with
only valence-like distributions at the initial $\mu^2$ scale and
relying on evolution to generate the distributions at higher $Q^2$. We
have evolved the predicted $\bar d(x) - \bar u(x)$ distribution from
$Q_0^2 = \mu^2 =0.25$ GeV$^2$ to $Q^2 = 54$ GeV$^2$. Since $\bar d(x)
- \bar u(x)$ is a flavor non-singlet parton distribution, its
evolution from $Q_0$ to $Q$ only depends on the values of $\bar d(x) -
\bar u(x)$ at $Q_0$, and is independent of any other parton
distributions. The solid curve in Fig.~\ref{fig_dbar-ubar} corresponds
to $\bar d(x) - \bar u(x)$ from the BHPS model evolved to $Q^2=$ 54
GeV$^2$. Significantly improved agreement with the data is now
obtained. This shows that the $x$-dependence of $\bar d(x) -\bar u(x)$
is quite well described by the five-quark Fock states in the BHPS
model provided that the $Q^2$-evolution is taken into
consideration. It is interesting to note that an excellent fit to the
data can be obtained if $\mu = 0.3$ GeV is chosen (dotted curve in
Fig.~\ref{fig_dbar-ubar}) rather than the more conventional value of
$\mu = 0.5$ GeV. We have also found good agreement between the HERMES
$\bar d(x) - \bar u(x)$ data at $Q^2 = 2.3 GeV^2$~\cite{hermes_sidis}
with calculation using the BHPS model.

\begin{figure}[t]
\includegraphics[width=0.5\textwidth]{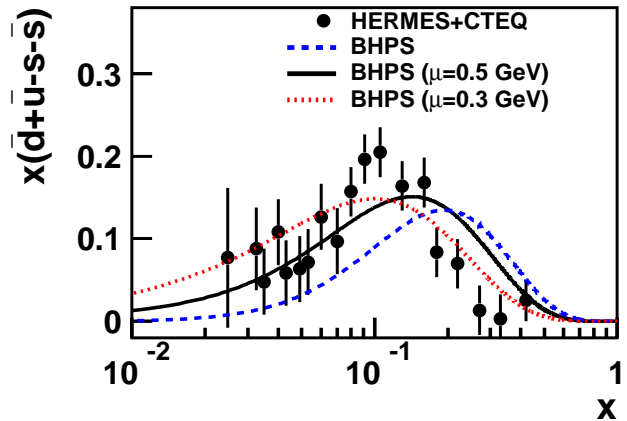}
\caption{Comparison of the $x(\bar d(x) + \bar u(x) - s(x) - \bar
s(x))$ data with the calculations based on the BHPS model. The dashed
curve corresponds to the calculation using Eq.~\ref{eq:prob5q_a}, and
the solid and dotted curves are obtained by evolving the BHPS result
to $Q^2 = 2.5$ GeV$^2$ using $\mu = 0.5$ GeV and $\mu = 0.3$ GeV,
respectively.}
\label{fig_sbar-dubar}
\end{figure}

We now consider the quantity $\bar u(x) + \bar d(x) - s(x) - \bar
s(x)$. New measurements of charged kaon production in semi-inclusive
DIS by the HERMES collaboration~\cite{hermes} allow the extraction of
$x(s(x) + \bar s(x))$ at $Q^2 = 2.5$ GeV$^2$. Combining this result
with the $x(\bar d(x) + \bar u(x))$ distributions determined by the
CTEQ group (CTEQ6.6)~\cite{cteq}, the quantity $x(\bar u(x) + \bar
d(x) - s(x) - \bar s(x))$ can be obtained and is shown in
Fig.~\ref{fig_sbar-dubar}. This approach for determining $x(\bar u(x)
+ \bar d(x) - s(x) - \bar s(x))$ is identical to that used by Chen,
Cao, and Signal in their recent study~\cite{signal} of strange quark
sea in the meson-cloud model~\cite{MCM2}.

An interesting property of $\bar u + \bar d - s - \bar s$ is that the
contribution from the extrinsic sea vanishes, just like the case for
$\bar d - \bar u$. Therefore, this quantity is only sensitive to the
intrinsic sea and can be compared with the calculation of the
intrinsic sea in the BHPS model. We have
\begin{eqnarray}
\bar u(x) + \bar d(x) - s(x) - \bar s(x) = \nonumber \\
P^{u \bar u}(x_{\bar u}) + 
P^{d \bar d}(x_{\bar d}) - 2 P^{s \bar s}(x_{\bar s}),
\label{eq:udssbar_p5}
\end{eqnarray}
\noindent where $P^{Q \bar Q}(x_{\bar Q})$ is the $x$-distribution for
$\bar Q$ in the $|u u d Q \bar Q\rangle$ Fock state.  Although the
shapes of the intrinsic $\bar u, \bar d, s, \bar s$ distributions can
be readily calculated from the BHPS model, the relative magnitude of
the intrinsic strange sea versus intrinsic non-strange sea is
unknown. We have adopted the assumption that the probability of the
intrinsic sea is proportional to $1/m_Q^2$, as stated earlier. This
implies that ${\cal P}^{s \bar s}_5/(\frac{1}{2}({\cal P}^{u \bar
u}_5 + {\cal P}^{d \bar d}_5)) = {m_{\bar u}^2}/{m_{\bar s}^2} \approx
0.36$ for $m_{\bar u} = 0.3$ GeV/c$^2$ and $m_{\bar s} = 0.5$
GeV/c$^2$. With this assumption, we can now compare the $x(\bar u(x) +
\bar d(x) - s(x) - \bar s(x))$ data with the calculation using the
BHPS model, shown as the dashed curve in
Fig.~\ref{fig_sbar-dubar}. The prediction of the BHPS model is found
to be shifted to larger $x$ relative to the data. This apparent
discrepancy could again partially reflect the different scales of the
theory and the data. Since $\bar u + \bar d - s - \bar s$ is a flavor
non-singlet quantity, we can readily evolve the BHPS prediction to
$Q^2 =2.5$ GeV$^2$ using $Q_0 = \mu = 0.5$ GeV and the result is shown
as the solid curve in Fig.~\ref{fig_sbar-dubar}. Better agreement
between the data and the calculation is achieved after the scale
dependence is taken into account. It is interesting to note that a
better fit to the data can again be obtained with $\mu = 0.3$ GeV,
shown as the dotted curve in Fig.~\ref{fig_sbar-dubar}.


From the comparison between the data and the BHPS calculation using
$\mu = 0.5$ GeV in Fig.~\ref{fig_sbar-dubar}, one can determine the
sum of the probabilities for the $|u u d u \bar u\rangle$ and $|u u d
d \bar d\rangle$ configurations, $\Sigma {\cal P}^{\bar d \bar u}_5$
($= {\cal P}^{d \bar d}_5 + {\cal P}^{u \bar u}_5)$. We found that
$\Sigma {\cal P}^{\bar d \bar u}_5 = 0.471$. Together with
Eq.~\ref{eq:intdbarubar2}, we have
\begin{equation}
{\cal P}^{u \bar u}_5 = 0.176;~~~~~{\cal P}^{d \bar d}_5 = 0.294.
\label{eq:ud_value}
\end{equation}
\noindent It is remarkable that the $\bar d(x) - \bar u(x)$ and the
$\bar d(x) + \bar u(x) - s(x) - \bar s(x)$ data not only allow us to
check the predicted $x$-dependence of the five-quark $|u u d u \bar
u\rangle$ and $|u u d d \bar d\rangle$ Fock states, but also provide a
determination of the probabilities for these two states. As expected,
the extracted values for the five-quark Fock states probabilities in
Eq.~\ref{eq:ud_value} depends on the assumption for the probability of
the $|uuds \bar s\rangle$. For the limiting case of ${\cal P}^{s \bar
s}_5=0$, we obtain ${\cal P}^{u \bar u}_5 = 0.097$ and ${\cal P}^{d
\bar d}_5 = 0.215$, which reflect the range of uncertainty of the
extracted values. It is interesting to note that values obtained in
Eq.~\ref{eq:ud_value} are consistent with the $1/m_Q^2$ assumption for
the probability of the $|u u d Q \bar Q\rangle$ Fock state. If one
uses the bag model estimate of ${\cal P}^{c \bar c}_5\sim
0.01$~\cite{donoghue77}, the $1/m_Q^2$ dependence would then imply
that ${\cal P}^{d \bar d}_5$ to be $\sim 0.01 (m_c^2/m_d^2) \sim
0.25$, consistent with the results of Eq.~\ref{eq:ud_value}.


In conclusion, we have generalized the existing BHPS model to the
light-quark sector and compared the calculation with the $\bar d -
\bar u$ and $\bar u + \bar d - s - \bar s$ data. The qualitative
agreement between the data and the calculation provides strong
supports for the existence of the intrinsic $u$ and $d$ quark sea and
the adequacy of the BHPS model. This analysis also led to the
determination of the probabilities for the five-quark Fock states for
the proton involving light quarks only. This result could guide future
experimental searches for the intrinsic $s$ and $c$ quark sea. This
analysis could also be readily extended to the hyperon and meson
sectors. The connection between the BHPS model and other multi-quark
models~\cite{zhang,bourrely} should also be investigated.

We acknowledge helpful discussion with Hai-Yang Cheng, Hung-Liang Lai,
Hsiang-Nan Li, and Keh-Fei Liu. This work was supported in part by the
National Science Council of the Republic of China and the
U.S. National Science Foundation. One of the authors (J.P.) thanks the
members of the Institute of Physics, Academia Sinica for their
hospitality.

\end{document}